\documentclass[aps,prc,reprint,groupedaddress]{revtex4-1}
\usepackage{epsfig}
\usepackage{multirow}
\begin{document}


\title{Breakdown of $N = 8$ magic number near the neutron drip line from parallel momentum distribution analyses }


\author{Shubhchintak}
\email{shub1dph@iitr.ac.in}
\author{R. Chatterjee}
\email{rcfphfph@iitr.ac.in}
\affiliation{Department of Physics, Indian Institute of Technology - Roorkee, 247667,
INDIA}


\date{\today}

\begin{abstract}
Using two theoretical models (the post form finite range distorted wave Born approximation and the adiabatic model) we calculate the parallel momentum distribution of the charged core in the Coulomb breakup of Be isotopes on a heavy target at 100 MeV/u. We show that the full width at half maxima of the parallel momentum distribution can be used to study the breakdown of $N = 8$ magic number away from the valley of stability.

\end{abstract}


\maketitle

{\it Introduction.}
The concept of magic numbers is one of the enduring features of the shell model. Based mostly on the study of nuclei on or close to the stability line it was found that if the neutron number ($N$) or the proton number (Z) was any of 2, 8, 20, 28, 50, 82 or 126, the nucleus was exceedingly stable. This was confirmed through the studies of Q-value of beta decay, the single nucleon separation energy, and excitation energy of the first excited state of an even-even nucleus \cite{bohr}. However, over the last two decades, the experiments performed with nuclei lying far off the stability line have shown that these magic numbers can disappear \cite{sorlin,gade} as one approaches the proton or neutron drip line. The breakdown of magic numbers is also related with the disappearance of the shell gaps or the mixing of single particle orbitals, which are well-separated in stable isotopes. For example, the halo structures of $^{11}$Be \cite{Talmi, Geithner} and $^{11}$Li \cite{Simon} can be explained by considering the valance neutron(s) in the intruder $2s_{1/2}$ orbit, that indicates the breakdown of $N = 8$ shell closure.

The breakdown of shell model magic numbers was first observed in neutron rich Na and Mg isotopes corresponding to $N = 20$ shell closure \cite{watt1}. In fact, in the region around $N = 20$, strongly deformed nuclei have been found where the inversion between normal-$sd$ and intruder-$pf$ shell has been suggested (the ``island of inversion") \cite{brown}. The large $B$(E2) values and low lying first excited states (which on the other hand for the nuclei at shell closures have been found at relatively large excitation energies), suggest the breakdown of magic number $N = 20$ \cite{detraz,moto}. 


The magic number $N = 8$ is the lowest magic number after the trivial $N = 2$ and it comes from any phenomenological potential model description of the nucleus, even without the spin-orbit coupling. The indications of $N = 8$ shell melting in neutron rich Be isotopes came from the abnormal ground state spin parity 1/2$^{+}$ observed in case of $^{11}$Be, which as already mentioned, was explained by considering the 2$s_{1/2}$ orbital lower in energy than the 1$p_{3/2}$ orbital. In case of $^{12}$Be, breakdown of $N = 8$ magic number has been suggested on the basis of its slow $\beta$-decay to  $^{12}$B \cite{suzuki}, and later on, this has been confirmed in many experiments \cite{Iwasaki,Navin,shimoura,Pain,Meharchand,Krieger}. As mentioned in Ref. \cite{Krieger}, the cluster structure formation in the low mass region could be a possible reason for shell melting. Another explanation, from a nuclear structure point of view, has been traced to the spin-isospin interaction between the proton and neutron orbit \cite{otsuka1, otsuka2, utsuno, sorlin}.


In this report we return to the study of magicity near the $N = 8$ (Be region) as one approaches the neutron drip line from a reaction point of view. The specific reaction observable that we choose is the parallel momentum distribution (PMD) of the charged fragment, in the Coulomb dissociation of the projectile in the field of a heavy target. Indeed it has been well known that the full width at half maxima (FWHM) of the PMD for the breakup of well known halo nuclei like $^{11}$Be and $^{19}$C is around 44 MeV/c, while that for stabler isotopes it is around over 140 MeV/c \cite{orr1,orr2}. Our hypothesis is that for the case of magic numbers a larger FWHM should be seen than the neighbouring isotopes.

{\it Formalism.}
We use two well established theoretical methods - the post form finite range distorted wave Born approximation (FRDWBA)~\cite{rc} and the adiabatic model (AD)~\cite{Tostevin}, for our calculations. Both theories are fully quantum mechanical and owe allegiance to the post form of the reaction theory. The initial and final state Coulomb interactions are also included to all orders.

We consider a breakup reaction of the type; $(a+t \rightarrow b+c+t)$, where the projectile $a$ break up into fragments $b$ (charged) and $c$ (uncharged) in the Coulomb field of a target $`t'$.

The PMD of the charged fragment $b$ is given by:
\begin{eqnarray}
\frac{d\sigma}{dp_{z}}&=&\int d\Omega_{c} dp_{x} dp_{y}m_{b}p_{b}\frac{2\pi}{\hbar v_{a}}\rho(E_{b},\Omega_{b},\Omega_{c})
\nonumber\\
&\times&\left\{\sum_{\ell m}\frac{1}{(2\ell+1)}|\beta_{\ell m}|^2\right\},
\end{eqnarray}
where $p_x$ and $p_y$ are the $x$- and $y$- components of the momentum $p_b$ of fragment $b$ having mass $m_b$. $v_a$ is the relative velocity of $a$ in the entrance channel and $\rho(E_{b},\Omega_{b},\Omega_{c})$ is the three-body final state phase space factor.

For the FRDWBA case, the reduced transition amplitude $\beta_{\ell m}$ is given by
\begin{eqnarray}
\beta_{\ell m}^{FRDWBA} &=& \left\langle e^{i(\gamma{\bf q_c} - \alpha{\bf K}).{\bf r_1}}\left|V_{bc}\right|\phi_{a}^{\ell m}({\bf r}_1)\right\rangle \nonumber\\
&\times&\left\langle \chi_{b}^{(-)}({\bf q_b},{\bf r_i})e^{i\delta {\bf q_c}.{\bf r_i}}|\chi_{a}^{(+)}({\bf q}_a,{\bf r_i})\right\rangle.
\end{eqnarray}

The ground state wave function of the projectile $\phi_{a}^{lm}({\bf r}_1)$ is contained in the first term (vertex function), while the second term which essentially describes the dynamics of the reaction, containing the Coulomb distorted waves $\chi^{(\pm)}$ can be expressed in terms of the bremsstrahlung integral ~\cite{Nordsieck}. $\alpha$, $\gamma$ and $\delta$ are mass factors pertaining to the three-body Jacobi coordinate system (see Fig. 1 of Ref. ~\cite{rc}). In Eq. 2, ${\bf K}$ is an effective local momentum appropriate to the core-target relative system and ${\bf q}_i$'s ($i = a, b, c$) are the Jacobi wave vectors of the respective particles. For more details on these quantities we refer to  Ref. \cite{rc}.

In case of the adiabatic approximation if one assumes that the dominant projectile breakup configurations excited are in the low energy continuum, then the reduced transition amplitude can again be written ~\cite{pb} into two parts - the structure part and the dynamics part, similar to Eq. (2) as, 
\begin{eqnarray}
\beta_{\ell m}^{AD} &=& \left\langle e^{i({\bf q_c} - \alpha{\bf q}_a).{\bf r_1}}\left|V_{bc}\right|\phi_{a}^{\ell m}({\bf r}_1)\right\rangle \nonumber\\
&\times& \left\langle \chi_{b}^{(-)}({\bf q_b},{\bf r})e^{i\delta {\bf q_c}.{\bf r}}|\chi_{a}^{(+)}({\bf q}_a,{\bf r})\right\rangle.
\end{eqnarray}

The input to this model as in Eq. (2) is again the full ground state wave function of the projectile $\phi_{a}^{lm}({\bf r}) = i^{\ell} u_{\ell}(r)Y_{\ell m}({\bf \hat r})$, where $u_{\ell}(r)$ is the radial part  and $Y_{\ell m}({\bf \hat r})$ are the spherical harmonics. To obtain a realistic $u_{\ell}(r)$, the radial Schr\"{o}dinger equation is solved with a Woods-Saxon potential (radius and diffuseness parameters taken as 1.15 fm and 0.5 fm respectively), whose depth is adjusted to reproduce the binding energy of the projectile. 

At this stage it is worthwhile to remember that although the breakup amplitudes,
Eqs. (2) and (3), look quite similar they are the result of different approximations
to the total wave function in the post form reaction theory. While the FRDWBA
formalism assumes that the breakup states are weakly coupled, the AD model
wavefunction is derived exactly if one makes the `adiabtic approximation',
stated above. Further discussion on these theories, including their applications on the breakup of halo nuclei on heavy targets can be found in Ref. \cite{rc}.

{\it Results and discussions.}
In Table \ref{tab:table1}, we present the FWHM from the PMD of the core in the Coulomb breakup of Be isotopes ($N = 5, 6, 7, 8$) on Au target at beam energy of 100 MeV/u, using both the FRDWBA and the adiabatic model.
\begin{table}[ht]
\caption{\label{tab:table1} FWHM from the PMD in the Coulomb breakup of Be isotopes on Au at 100 MeV/u beam energy. Shown also are the ground state spin-parities (J$^\pi$), ground state single particle configurations, one neutron separation energies ($S_{n}$) \cite{nndc1} of the various Be isotopes considered.
 Note that the FWHM for the breakup of $^{10}$Be ($N = 6$) is the highest, rather than $^{12}$Be ($N = 8$), having the magic number of neutrons.
 }
\begin{tabular}{|c|c|c|c|c|*{2}{c|}}
\hline 
Proj- & $N$ &(J$^\pi$) &single particle & $S_{n}$ & \multicolumn{2}{|c|}{FWHM} \\
ectile&& &state& (MeV) & \multicolumn{2}{|c|}{(MeV/c)}\\
\hline
 & & & & & FRDWBA & AD \\
\hline
 $^{9}$Be& 5 & 3/2$^-$ & $^{8}$Be$(0^{+})$$\otimes$${1p_{3/2}}$$\nu$  & 1.665 & 112.27 & 113.87 \\
$^{10}$Be& 6 & 0$^+$   & $^{9}$Be$(3/2^{-})$$\otimes$${1p_{3/2}}$$\nu$& 6.812 & 191.13 & 170.30 \\
$^{11}$Be& 7 & 1/2$^+$ &$^{10}$Be$(0^{+})$$\otimes$${2s_{1/2}}$$\nu$  & 0.501 & 43.23  & 43.71 \\
$^{12}$Be& 8 & 0$^+$   &$^{11}$Be$(1/2^{+})$$\otimes$${2s_{1/2}}$$\nu$& 3.169 & 88.93  & 89.73  \\

\hline
\end{tabular}
\end{table}
 
Let us now make a few comments on the single particle structure of the Be isotopes considered in Table \ref{tab:table1}. It is clear from the table that the ground state spin-parity J$^\pi$ (3/2$^-$) of $^{9}$Be is obtained according to the shell model, where we consider the coupling of $p_{3/2}$ neutron with $^{8}$Be$(0^{+})$ core, having threshold energy 1.665 MeV. Interestingly, addition of one more neutron to $^{9}$Be leads to a tightly bound $^{10}$Be nucleus having $^{9}$Be + n separation energy 6.812 MeV. This is an even-even nucleus having J$^\pi = 0^+$ and also follow the normal shell ordering. However, further addition of one more neutron leads to $^{11}$Be, which is one of the oldest example of intruder configurations \cite{Talmi}, where 2$s_{1/2}$ orbital is situated below the 1$p_{1/2}$ orbital. This is a well known one-neutron halo nucleus with $S_n = 0.501$ MeV.

The next isotope is $^{12}$Be, which corresponds to $N = 8$ shell closure. Its ground state spin suggests that the possible configuration of the last two valance neutrons could be $(p_{1/2})^2$, $(s_{1/2})^2$ or $(d_{5/2})^2$. However, it has been shown in many theoretical \cite{Barker1,Fortune,Nune, Romero} as well as experimental studies \cite{Iwasaki,Navin,shimoura,Pain,Meharchand} that there is an admixture of 2$s$1$d$ and 1$p$ orbitals in the ground state of $^{12}$Be, which is also an indication of the breakdown of $N = 8$ magic number. However, for our calculations we take the dominant single particle configuration, $^{11}$Be$(1/2^{+})$$\otimes$${2s_{1/2}}$$\nu$, as in Ref. \cite{Navin}.
The N = 9 Be isotope ($^{13}$Be) is known to have only two levels \cite{kondonaka}, both of which are resonance states with some uncertainties in their positions. Keeping this aside, we shall therefore limit our analyses from
$N=5-8$ Be isotopes.

\begin{figure}[h]
\centering
\includegraphics[height=6.0cm, clip,width=0.45\textwidth]{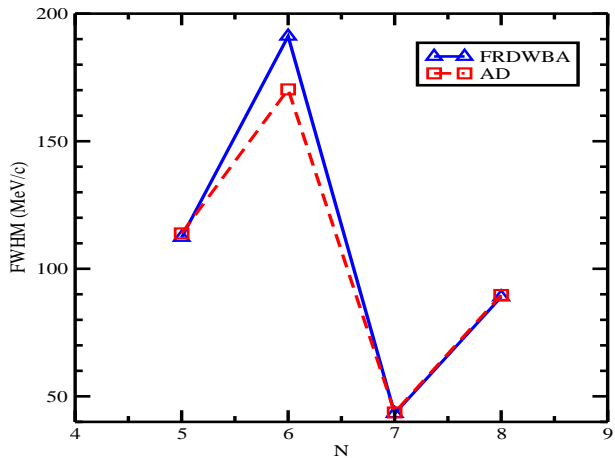}
\caption{\label{fig1} (Color online) FWHM of PMD of charged fragment, calculated in the Coulomb breakup of various Be isotopes plotted with respect to neutron number $N$. The triangles and squares are results of FRDWBA and AD calculations, respectively. The lines (solid and dashed) are a guide to the eye. Among Be isotopes $^{10}$Be ($N = 6$) has the highest FWHM.}
\end{figure}

We now turn our attention back to the FWHM calculated in Table \ref{tab:table1}. The FWHM of the PMD is the smallest for $^{10}$Be, which is obtained from the breakup of  the well known halo nucleus $^{11}$Be. Our calculated results compare quite well with the experimental value of $43.6 \pm 1.1$ MeV/c \cite{orr1}, obtained from the breakup of $^{11}$Be on heavy targets at 63 MeV/u beam energy.
For the case of $N = 8$, experimental results exist for the breakup of $^{12}$Be
on a light target \cite{Navin} at 78 MeV/u beam energy. Their value of 89 MeV/c (extracted from Fig. 2(a) of Ref. \cite{Navin}, for the $s$-state) compares quite well
with our breakup calculations on a heavy target. 

At this stage it is worthwhile to note that the width of the PMD is independent of the reaction mechanism. In fact, two empirical models of fragmentation by Goldhaber \cite{gold} and Morrissey \cite{morri}, suggest that the width of PMD does not depend upon target mass at all. This has also been confirmed in many experiments \cite{orr2,orr3,bazin1} as well as in theoretical studies \cite{bertulani3,banerjee} involving fragmentation reactions. Next comes the question of beam energy dependence of PMD in the breakup process. It has been observed that the width of the PMD remains nearly constant for a wide range of beam energies (50 MeV/u - 2 GeV/u) \cite{kidd,michel}, except for those below 10 MeV/u \cite{friedman,ck}. However, it is important to note that for an analysis of the type reported in our study, calculations or experiments must be done in the same range of beam energies, in order to make any inference independent of beam energy.   

Nevertheless, it is interesting to note that the maximum FWHM is obtained for $N = 6$ and not for $N = 8$ (the usual magic number). This is indeed a comment on the breakdown of magicity for $N = 8$ near the drip line. 
However, to dispel any ambiguity, let us also analyse the effect of mixing various single particle configurations in accordance with their spectroscopic factors (0.42 for $s$-state and 0.37 for $p$-state) in $^{12}$Be \cite{Navin}. The $p$-state, which couples with the $1/2^-$ excited state (0.32 MeV above the ground state \cite{nndc1}) of $^{11}$Be corresponds to the configuration $^{11}$Be$(1/2^{-})$$\otimes$${1p_{1/2}}$$\nu$. With this mixing (including proper spectroscopic factors) the FWHM for the PMD turns out to be 100.50 MeV/c and 101.94 MeV/c for FRDWBA and AD cases, respectively. This is hardly a ten percent change in the width of the PMD and is no way near the N = 6 case ($^{10}$Be).
We do not expect any significant contribution to the width by including the higher angular momentum $d$-state which has to be coupled with $5/2^+$ resonance state (1.78 MeV above the ground state and width 100 keV \cite{nndc1}) of $^{11}$Be corresponding to the configuration $^{11}$Be$(5/2^{+})$$\otimes$${1d_{5/2}}$$\nu$. Therefore, we continue our analysis with the predominant $s$-wave configuration of  $^{12}$Be. 

For a more clear view, in Fig. \ref{fig1}, we have plotted the calculated FWHM with respect to $N$. The triangles are the results of FRDWBA calculations, whereas the square boxes are the results obtained from AD model.
Let us now return to the central hypothesis of this study. Halo nuclei, which
are weakly bound, have a narrow PMD. However if the isotope under consideration
has a magic number of neutrons, it is \textit{supposed} be stabler than its neighbouring counterparts. We find it interesting that in the Be chain $^{12}$Be (N = 8) breakup does not have the largest FWHM. Rather the largest value of FWHM obtained corresponding to $N = 6$ (case of $^{10}$Be) suggests that $N = 6$ could be a magic number, which is also in agreement with the studies of Refs. \cite{otsuka1, sorlin}.

{\it Conclusions.}
In conclusion, using Coulomb breakup reactions on a heavy target we studied the breakdown of magic number $N = 8$ for the neutron rich Be isotopes using PMD of the charged fragment. A relatively small FWHM of PMD of the core in the breakup of $^{12}$Be is an indication of the breakdown of $N = 8$ magic number. In contrast, $N = 6$ shows the signature of a magic number. The width of the PMD is known to be independent of the target mass in the fragmentation process and is also nearly constant over a wide range of beam energies. Therefore, our hypothesis to use the PMD in Coulomb breakup to study magic numbers can be used to predict breakdown and emergence of new magic numbers in exotic nuclei in a simple manner. Further applications of this hypothesis to study other magic numbers and shell gaps are in progress. 
\\

This text presents results from research supported by the Department of Science and Technology, Govt. of India (SR/S2/HEP-040/2012).



\begin{thebibliography}{99}
\bibitem{bohr} A. Bohr and B. R. Mottelson, Nuclear Structure Vol. 1, World Scientific Publ. Co. Ltd., 1998.
\bibitem{sorlin} O. Sorlin and M.-G. Porquet, Prog. Part. Nucl. Phys. {\bf 61}, 602 (2008).
\bibitem{gade} A. Gade and T. Glasmacher, Prog. part. nucl. phys. {\bf60}, 161 (2008).
\bibitem{Talmi} I. Talmi and I. Unna, Phys. Rev. Lett. {\bf 4}, 469 (1960).
\bibitem{Geithner} W. Geithner {\it et al.}, Phys. Rev. Lett. {\bf 83}, 3792 (1999).
\bibitem{Simon} H. Simon {\it et al.}, Phys. Rev. Lett. {\bf 83}, 496 (1999).
\bibitem{watt1} A. Watt, R. P. Singhal, M. H. Storm and R. R. Whitehead, J. Phys. G {\bf 7}, L145 (1981); M. H. Storm, A. Watt and R. R. Whitehead, J. Phys. G {\bf 9}, L165 (1983). 
\bibitem{brown} E.K. Warburton, J.A. Becker, B.A. Brown, Phys. Rev. C {\bf 41}, 1147 (1990).
\bibitem{detraz} C. D{\'e}traz {\it et al.}, Phys. Rev. C {\bf 19}, 164 (1979).
\bibitem{moto} T. Motobayashi {\it et al.}, Phys. Lett. B {\bf 346}, 9 (1995).
\bibitem{suzuki} T. Suzuki and T. Otsuka, Phys. Rev. C {\bf 56}, 847 (1997).
\bibitem{Iwasaki} H. Iwasaki {\it et al.}, Phys. Lett. B {\bf481}, 7 (2000); H. Iwasaki {\it et al.}, Phys. Lett. B {\bf491}, 8 (2000).
\bibitem{Navin} A. Navin {\it et al.}, Phys. Rev. Lett. {\bf 85}, 266 (2000).
\bibitem{shimoura} S. Shimoura et al., Phys. Lett. B {\bf 560}, 31 (2003).
\bibitem{Pain} S. D. Pain {\it et al.}, Phys. Rev. Lett. {\bf 96}, 032502 (2006).
\bibitem{Meharchand} R. Meharchand {\it et al.}, Phys. Rev. Lett. {\bf 108}, 122501 (2012).
\bibitem{Krieger} A. Krieger {\it et al.}, Phys. Rev. Lett. {\bf 108}, 142501 (2012).
\bibitem{otsuka1} T. Otsuka, R. Fujimoto, Y. Utsuno, B. A. Brown, M. Honma, and T. Mizusaki, Phys. Rev. Lett. {\bf 87}, 082502 (2001)
\bibitem{otsuka2} T. Otsuka, T. Suzuki, R. Fujimoto, H. Grawe, and Y. Akaishi, Phys. Rev. Lett. {\bf95}, 232502 (2005).
\bibitem{utsuno} Y. Utsuno, T. Otsuka, T. Mizusaki and M. Honma, J. Phys: Conf. Series {\bf 49}, 126 (2006).


\bibitem{orr1} J. H. Kelley {\it et al.}, Phys. Rev. Lett. {\bf 74}, 30 (1995).
\bibitem{orr2} E. Sauvan {\it et al.}, Phys. Rev C {\bf 69}, 044603 (2004).
\bibitem{rc} R. Chatterjee, P. Banerjee and R. Shyam, Nucl. Phys. A {\bf675}, 477 (2000).
\bibitem{Tostevin} J. A. Tostevin et al., Phys. Lett. B {\bf 424}, 219 (1998). 
\bibitem{Nordsieck} A. Nordsieck, Phys. Rev. {\bf 93}, 785 (1954).
\bibitem{pb} P. Banerjee, I. J. Thompson and J. A. Tostevin, Phys. Rev. C {\bf58}, 1042 (1998).
\bibitem{nndc1} {\it http://www.nndc.bnl.gov/nudat2}.
\bibitem{Barker1} F. C. Barker, J. Phys. G {\bf 2}, L45 (1976); F. C. Barker, J. Phys. G {\bf 36}, 38001 (2009).
\bibitem{Fortune} H. T. Fortune and R. Sherr, Phys. Rev. C {\bf 83}, 044313 (2011).
\bibitem{Nune} F. Nunes, I. J. Thompson and J. A. Tostevin, Nucl. Phys. A {\bf 703}, 593 (2002).
\bibitem{Romero} C. Romero-Redondo and E. Garrido and D. V. Fedorov and A. S. Jensen, Phys. Rev. C {\bf 77}, 054313 (2008).
\bibitem{kondonaka} Y. Kondo {\it et al.}, Phys. Lett. B {\bf 690}, 245 (2010).
\bibitem{gold} A. S. Goldhaber, Phys. Lett. B {\bf 53}, 306 (1974).
\bibitem{morri} D. J. Morrissey, Phys. Rev. C {\bf 39}, 460 (1989).
\bibitem{orr3} N. A. Orr {\it et al.}, Phys. Rev. Lett. {\bf 69}, 2050 (1992); N. A. Orr {\it et al.}, Phys. Rev. C {\bf 51}, 3116 (1995). 
\bibitem{bazin1} K. Meierbachtol, D. J. Morrissey, M. Mosby and D. Bazin, Phys. Rev. C {\bf 85}, 034608 (2012)
\bibitem{bertulani3} C. A. Bertulani and K. W. McVoy, Phys. Rev. C {\bf 46}, 2638 (1992).
\bibitem{banerjee} P. Banerjee and R. Shyam, Phys. Lett. B {\bf 349}, 421 (1995).
\bibitem{michel} M. C. Mermaz, Phys. Rev C {\bf 36}, 1000 (1987).
\bibitem{kidd} J. M. Kidd, P. J. Lindstrom, H. J. Crawford and G. Woods, Phys. Rev. C {\bf 37}, 2613 (1988).
\bibitem{ck} C.K. Gelbke {\it et al.}, Phys. Lett. B {\bf 70}, 415 (1977).
\bibitem{friedman} W. A. Friedman, Phys. Rev. C {\bf 27}, 569 (1983).



\end{thebibliography}

\end{document}